**Effect of Mg diffusion on bilayer photoluminescence spectra of $Mg_{0.29}Zn_{0.71}O$/ZnO interface at different annealing temperatures**

Amit K. Das*[1]

[1]*Laser Material Proccessing Division, Raja Ramanna Centre for Advanced Technology, Indore, India – 452013*

*Corresponding author's email : amitkrdh@gmail.com

Tel: +91 731 248 8304; Fax: +91 731 248 8300

**Abstract:**

$Mg_{0.29}Zn_{0.71}O$/ZnO (MZO/ZnO) bilayers were grown on sapphire substrates by pulsed laser deposition at ~ $400^0C$ and were subsequently annealed at different temperatures ranging from ~ 500 to $900^0C$ to allow diffusion of Mg across the interface. The diffusion of Mg was confirmed from secondary ion mass spectroscopy (SIMS) and Photo-absorption measurements on the samples. As a result of Mg diffusion the photoluminescence (PL) spectra of the bilayers got significantly modified. Whereas the as grown structure had two distinct near band edge UV luminescence peaks corresponding to the ZnO and MZO layer, the one annealed at ~ $900^0C$ had a single near band edge PL peak corresponding to a complete intermixed layer. At the intermediate annealing temperature of ~$800^0C$ the PL peak position corresponding to the MZO layer red-shifted while that corresponding to the ZnO layer blue-shifted, implying a diffused interface and partial intermixing. A theoretical model has been developed and numerically simulated to correlate the PL spectra of the bilayers with the Mg diffusion profile across the interface, as measured by SIMS.

**Keywords:** ZnO/MgZnO bilayer, PLD, Photoluminescence, Mg diffusion

# 1. Introduction:

MgZnO/ZnO heterostructures are important for confining electrons in ZnO for various applications like quantum well lasers, resonant-tunneling diodes, high electron mobility transistors, quantum Hall effects etc[1- 4]. In all these applications it is a primary requirement that the heterointerface between MgZnO and ZnO layers be of high quality to obtain higher carrier mobility, narrow and stronger excitonic transitions lines and to enhance the overall device performance. Generally high growth temperatures in the range of ~ 600 to $800^0$C are used for the growth of good quality MgZnO thin films by different deposition methods [1, 4, 5]. Moreover the solubility limit of MgO in MgO-ZnO binary system is reported to be ~ 4 mol% under thermal equilibrium condition [6]. But for practical applications much higher Mg concentrations (than the solubility limit) are needed which essentially leads to the formation of a metastable phase in MgO-ZnO alloys [6]. These two facts, namely- the high growth temperature and metastable growth mode may have serious consequences on the quality of interface in MgZnO/ZnO bilayers due to Mg diffusion across the interface during growth or during device performance. This diffusion of Mg across the interface can significantly alter the photoluminescence (PL) spectra of the MgZnO/ZnO quantum wells. In this context we studied the photoluminescence of $Mg_{0.29}Zn_{0.71}O$/ZnO bilayer systems under different conditions of annealing to ascertain the effects of Mg diffusion. A theoretical model has also been developed and numerically simulated to correlate the photoluminescence spectra from the bilayer with the diffusion of Mg across the interface. The result of these experiments and theoretical calculations are presented here.

## 2. Materials and methods:

The $Mg_{0.29}Zn_{0.71}O$/ZnO heterostructures were grown on c-axis (0001) sapphire substrates using pulsed laser deposition (PLD) with a KrF excimer laser. The bottom ZnO layer was grown at the substrate temperature of ~ $750^0C$ and then the top $Mg_{0.29}Zn_{0.71}O$ layer were grown at $400^0C$. The oxygen partial pressure in the PLD chamber was maintained at ~ $10^{-5}$ mbar during the growth. A set of $Mg_xZn_{1-x}O$ thin films with varying *x* were also grown on sapphire substrates using PLD under similar conditions to calibrate the photoluminescence (PL) peak position of MZO with Mg composition as required for our theoretical calculations. The successful growth of the bilayer system was confirmed from absorption spectra and secondary ion mass spectroscopy measurements (SIMS). SIMS was used to measure the Mg composition as well as the depth profile of Mg distribution across the thickness of the films. Photoluminescence (PL) spectra of the heterostructures/films were recorded using a spectrometer with an excitation laser at 266 nm.

## 3. Result and Discussion:

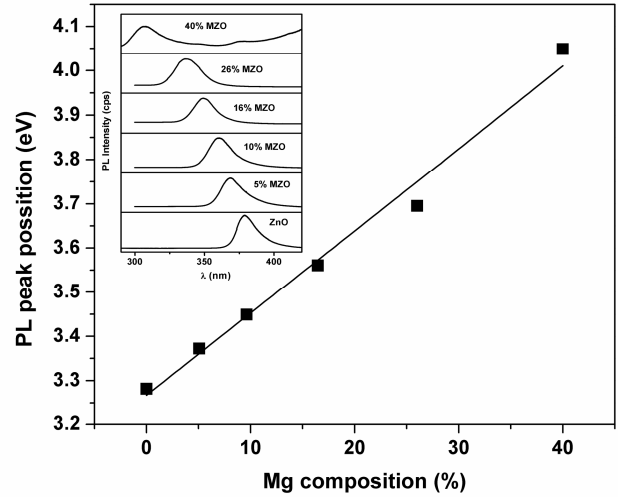

**Figure 1**

Figure 1 shows the PL peak positions of the $Mg_xZn_{1-x}O$ films as a function of $x$ in the range from 0 to ~ 40% whereas the inset shows the corresponding PL spectra of the films. It can be seen that the PL peak positions are having a linear relationship with the Mg concentration ($x$) in the films. This is in conformity with the available literature [7]. The linear fit of the data could be well described by the equation,

$$E_{PL}(x) = E_{PL}(0) + 0.019x \qquad (1)$$

Where $E_{PL}$ is the PL peak position in eV of $Mg_xZn_{1-x}O$ film and $E_{PL}(0)$ is 3.29 eV. This equation will be used later in our theoretical calculations.

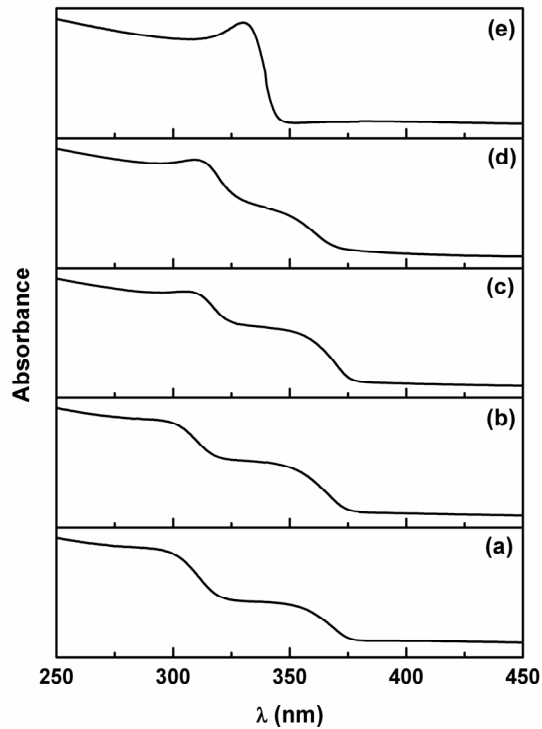

**Figure 2**

Figure 2 shows the absorption spectra of the as grown and annealed $Mg_{0.29}Zn_{0.71}O/ZnO$ bilayers. The absorption spectrum of the as grown heterostructure showed quite sharp absorption edges for the ZnO and the MZO layers at ~ 372 and 310 nm respectively. This confirmed the successful growth of MZO/ZO heterostructure with well defined interface under the used experimental conditions. As the heterostructures were subsequently annealed for three hours at temperatures from 600 to $900^0C$, it was observed that the bilayer maintained the positions and sharpness of the absorption edges till the annealing temperature of $700^0C$ implying no significant Mg diffusion across the interface. However, at the annealing temperature of $800^0C$, the absorption edge of ZnO layer blue-shifted and that of the MZO layer red-shifted as a result of Mg diffusion. As the annealing temperature was raised to $900^0C$ the two layers mixed completely with each other as was apparent from the single absorption edge.

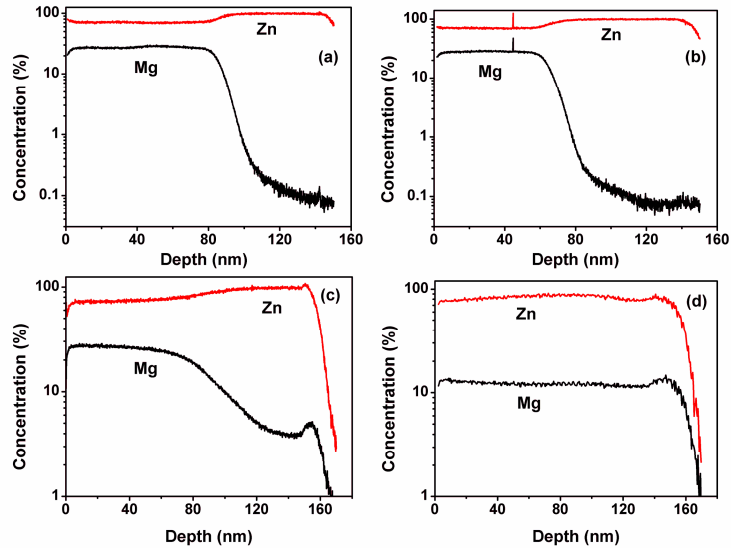

**Figure 3**

The SIMS depth profiles for the as grown and annealed bilayers are shown in figure 3. For the as grown bilayer the interface was quite sharp and the Mg concentration decreased from ~29% to less than 1% within a depth of ~ 20 nm at the interface as shown in figure 3(a). The bilayer that was annealed at $700^0C$ also showed similar Mg distribution with depth as depicted in figure 3(b). These results are in conformity with the absorption spectra of the samples. However, at the annealing temperature of $800^0C$, significant Mg diffusion into the ZnO layer was apparent as can be seen in figure 3(c). The Mg concentration decreased gradually from ~ 25% to ~ 4% over depth of ~ 100 nm, thereby rendering the interface highly diffused. Mg was present in the entire depth of the ZnO layer thereby blue-shifting its band edge. Also as a result of diffusion, the Mg composition in the MZO layer decreased. When the annealing temperature was further increased to $900^0C$, the bilayer got reduced to a single MZO layer with ~ 13% Mg as shown in figure 3(d). The single absorption edge seen in the absorption spectrum of this sample was therefore in conformity with the SIMS depth profile.

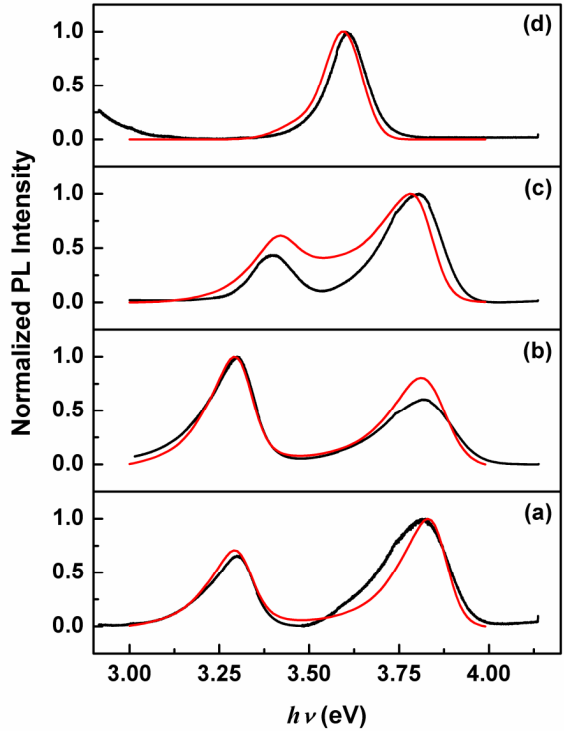

**Figure 4**

The photoluminescence (PL) spectra of the as grown and annealed $Mg_{0.29}Zn_{0.71}O/ZnO$ heterostructures are shown in figure 4. The as grown heterostructure showed two distinct peaks corresponding to the ZnO and MZO layers at ~3.30 and ~3.82 eV respectively. The peak positions are maintained at the same positions at the annealing temperature of $700^0C$ as well. These were expected since there were no significant diffusion of Mg and ZnO across the interface as proved by SIMS depth profiles and absorption spectra for these samples. However as the annealing temperature was raised to $800^0C$, the PL peak corresponding to ZnO layer blue-shifted to ~ 3.40 eV while that corresponding to the

MZO layer slightly red shifted to ~ 3.80 eV as a result of Mg diffusion across the interface. For the heterostructure annealed at 900$^0$C, only a single PL peak could be observed at ~ 3.60 eV corresponding to a single MZO layer without any interface, as expected.

The observed PL spectra can be theoretically correlated with the experimental Mg depth profiles measured by SIMS by the following method. The SIMS depth profiles gave Mg concentration $x(y)$ as a function of depth $y$ from surface. As described earlier the Mg concentration can be related to the PL peak position $E_{PL}(x)$ using equation (1). We can divide the whole depth profile into many infinitesimally small elements $dy$ at a distance $y$ from the surface with constant Mg composition, each of which will give PL corresponding to its Mg concentration with peak positions given by equation (1). The contribution to the overall PL at the frequency $\nu$ due to this small element will be given by,

$$dI(\nu) = G(\nu, E_{PL})dy \qquad (2)$$

Where $G(\nu, E_{PL})$ is the PL line shape function for the element with PL peak position at $E_{PL}$. We approximated it, for our calculations, as the PL profile of pure ZnO thin films which was found by fitting actual PL profile data of ZnO by three Gaussians and is given by,

$$G(\nu, E_{PL}) = \sum_{k=1,2,3} A_k \exp\left(-\frac{2(h\nu - E_{PL} + \Delta_k)^2}{\sigma_k^2}\right) \qquad (3)$$

Where $\sigma_1 = .074$ eV, $\sigma_2 = 0.12$ eV, $\sigma_3 = 0.21$ eV, $A_1 = 2.4$, $A_2 = 2.6$, $A_3 = 1.0$, $\Delta_1 = 0$, $\Delta_2 = 0.042$ eV and $\Delta_3 = 0.11$ eV. The line shape function $G(\nu, E_{PL})$ is a function of $y$ via $E_{PL}$. The overall PL profile then will be given by,

$$I(\nu) = \int_0^t G(\nu, E_{PL}) dy \quad (4),$$

where $t$ is the total thickness of the sample. The integration was evaluated numerically using GNU scientific library taking the SIMS depth profile data $x(y)$ as the input. The resultant PL spectra are plotted on figure 4 in red. As can be seen it agreed reasonably well with the experimental PL spectra. The discrepancy that was there between the experimental and theoretical PL spectra was plausibly due to the use of some simplifying assumptions. The form of the line shape function and the PL efficiency may actually depend on Mg concentration and annealing temperature which was ignored in the calculations.

## 4. Conclusion

In conclusion we have successfully grown $Mg_{0.29}Zn_{0.71}O$/ZnO (MZO/ZnO) bilayer by pulsed laser deposition on sapphire substrates at $400^0C$. The bilayers were subsequently annealed at different temperatures from 500 to $900^0C$ to allow the diffusion of Mg across the interface. Secondary ion mass spectroscopy (SIMS) and Photo-absorption on the samples confirmed Mg diffusion. This intermixing of the two layers at the interface modified the photoluminescence (PL) from the heterostructures significantly. The modified PL spectra from the bilayers have been correlated with Mg diffusion across the interface by developing a theoretical model and numerically simulating it.

**Figure Captions:**

**Figure 1.** Variation of PL peak positions of the $Mg_xZn_{1-x}O$ films as a function of *x*. The inset shows the corresponding PL spectra of the films.

**Figure 2.** The absorption spectra of the (a) as grown and annealed $Mg_{0.29}Zn_{0.71}O/ZnO$ bilayers at (b) $600^0C$, (c) $700^0C$, (d) $800^0C$ and (e) $900^0C$.

**Figure 3.** Variation of Mg and Zn concentration (measured by SIMS) as a function of depth for the (a) as grown and annealed $Mg_{0.29}Zn_{0.71}O/ZnO$ bilayers at (b) $700^0C$, (c) $800^0C$ and (d) $900^0C$.

**Figure 4.** The experimental (black) and theoretical (red) photoluminescence spectra of the (a) as grown and annealed $Mg_{0.29}Zn_{0.71}O/ZnO$ bilayers at (b) $700^0C$, (c) $900^0C$ and (d) $900^0C$.